\documentclass[prd,aps,amsmath,amssymb,nofootinbib]{revtex4}
\usepackage{amsmath}
\usepackage{graphicx}

\def\half{{1\over 2}}

\def\foot{\footnote}
\def\be{\begin{equation}}
\def\ee{\end{equation}}

\begin{document}

\title{Dark Energy: a Brief Review}

\author{Miao Li$^{1,2}$, Xiao-Dong Li$^{2,3,4}$, Shuang Wang$^{5,6}$, Yi Wang$^{7,8}$}

\affiliation{1) Institute of Theoretical Physics, Chinese Academy of Sciences, Beijing 100190, China}
\affiliation{2) Kavli Institute for Theoretical Physics, Key Laboratory of Frontiers in Theoretical Physics, Beijing 100190, China}
\affiliation{3) Department of Modern Physics, University of Science and Technology of China, Heifei, 230026, China}
\affiliation{4) Interdisciplinary Center for Theoretical Study, University of Science and Technology of China, Hefei 230026, China}
\affiliation{5) Department of Physics, College of Sciences, Northeastern University, Shenyang 110004, China}
\affiliation{6) Homer L. Dodge Department of Physics \& Astronomy, Univ. of Oklahoma, 440 W Brooks St., Norman, OK 73019, U.S.A.}
\affiliation{7) Kavli Institute for the Physics and Mathematics of the Universe, Todai Institutes for Advanced Study, University of Tokyo, 5-1-5 Kashiwanoha, Kashiwa, Chiba 277-8583, Japan}
\affiliation{8) Department of Physics, McGill University, Montr\'eal, QC, H3A 2T8, Canada}

\begin{abstract}
The problem of dark energy is briefly reviewed in both theoretical and observational aspects. In the theoretical aspect, dark energy scenarios are classified into symmetry, anthropic principle, tuning mechanism, modified gravity, quantum cosmology, holographic principle, back-reaction and phenomenological types. In the observational aspect, we introduce cosmic probes, dark energy related projects, observational constraints on theoretical models and model independent reconstructions.
\end{abstract}

\maketitle

\section{Introduction}

The Universe probed by gravitational methods looks very different from that appears optically.
Thanks to a large number of observational windows on the gravitational side,
the difference can be classified and explained by two additional components in our Universe:
a component behaves like cold particles dubbed cold dark matter, and a component without significant dynamics dubbed dark energy.
Both components bring big challenges to the current understanding of physics.

Here we focus on dark energy \cite{DEReviews}.
Whereas observational evidences of dark energy emerges within the past 15 years,
theoretical considerations of dark energy began almost a century ago. Those theoretical considerations are known as the ``cosmological constant problem''.

In 1917, the early days of general relativity, Einstein \cite{EinsteinCE} added a cosmological constant to his field equations to obtain a static Universe
\footnote{He removed this term from his equations later in 1931, due to the discovery of cosmic expansion. }.
Soon in the 1920s, Pauli realized that the quantum zero point energy of a radiation field is too large to gravitate.
Pauli's statement is already surprisingly close to the modern understanding of the cosmological constant problem,
considering the fact that now we understand fundamental physics much better than a century ago.

After that, the cosmological constant problem was largely ignored (or it was too hard to push forward) for a long time until the 1960s.
In 1967, Zeldovich considered the cosmological constant problem as a fine tuning problem \cite{ZeldovichGD}.
This is known as the old cosmological constant problem.
Since then, a lot of early efforts have been made to eliminate the too large cosmological constant, as reviewed in \cite{weinrev}.

In 1998, supernovae observations eventually shed light on this hard problem \cite{riess}\cite{perl},
and in an unexpected way. Data suggests that dark energy, an energy component probably coming from the cosmological constant,
dominates the current Universe. The percentage of dark energy density is around 70\% -- not small nor too large.
The Universe is now accelerating due to the presence of dark energy.

In this essay we review the problem of dark energy, a.k.a. the cosmological constant problem.
We first review the problem in the viewpoint of theoretical physics in logic order, from the aspects of symmetry, anthropic principle,
tuning mechanisms, modified gravity, quantum cosmology, holographic principle, back-reaction and phenomenological models.
Then we review the observational topics, including cosmic probes, dark energy related projects,
observational constraints on theoretical models and model independent reconstructions.

\section{Theoretical aspects}

The theoretical challenge for dark energy can be observed from the Einstein equations
\begin{align}
\label{eins}
R_{\mu\nu}-\half g_{\mu\nu}R+\Lambda g_{\mu\nu}=8\pi GT_{\mu\nu},
\end{align}where $\Lambda$ is the bare cosmological constant, as an input parameter from the theory.

On the right hand side, the energy-momentum tensor $T_{\mu\nu}$ is to be determined by a given matter configuration. Note that a perturbative matter state is built up from the vacuum state. Also our Universe looks not very far from the vacuum state. Thus to see the problem, it suffices to consider a vacuum stress tensor.

For the vacuum, Lorentz invariance forces the stress tensor to take the form
\begin{align}
\label{TCemtFluid}
 T_{\mu\nu}=(\rho_{\mathrm{vac}}
+p_{\mathrm{vac}} )u_\mu u_\nu +p_{\mathrm{vac}} g_{\mu\nu} = -\rho_{\mathrm{vac}} g_{\mu\nu},
\end{align}
where $p_{\mathrm{vac}} = -\rho_{\mathrm{vac}}$. Thus $\rho_{\mathrm{vac}}$ behaves as a renormalization of $\Lambda$. The quantum correction to the vacuum energy density can be calculated as (taking a bosonic field as an example)
\begin{align}
\label{zeropint}
\rho_{\mathrm{vac}}=\frac{1}{2V}\sum_i\omega_i= \int{k^2dk\over 4\pi^2}(k^2+m^2)^\half
\simeq {\lambda^4\over 16\pi^2},
\end{align}
where $V$ is the volume of space and in the last step we have used a UV cutoff $\lambda$ to evaluate the divergent integral. If an effective field theory were applicable up to the energy scale $\lambda$, and natural, we could expect the following facts:

(1) The energy density $\rho_{\mathrm{vac}}$ gravitates. This is hard to avoid in a local field theory because the current test of equivalence principle has been far more precise than the quantum contribution to the mass of matter. And a local graviton cannot observe a difference between a vacuum loop and a loop diagram with external legs.

(2) The effective field theory contribution to energy density $\rho_{\mathrm{vac}}$ is not canceled by the high energy modes above the cutoff. This is again hard to avoid unless there are some nontrivial (and yet unknown) UV-IR correlations.

(3) The energy density $\rho_{\mathrm{vac}}$ is not canceled by contribution from other fields. If there were supersymmetry (SUSY), the bosonic vacuum energy is actually canceled by its fermionic counterpart. However SUSY, if exists, is broken on energy scales below $100$GeV. Thus below $100$GeV there is no known cancellation between fields.

(4) The energy density $\rho_{\mathrm{vac}}$, together with contributions from other fields, are not canceled by the bare energy cosmological constant $\Lambda$. Those two do not seem to be able to cancel each other unless gravity is modified to fit matter in a more dynamical way.

Given the above ``facts'', our observed vacuum energy density should be of order $\rho_{\mathrm{vac}}$, which is $10^{71}$GeV$^4$ if $\lambda=M_p$, or $10^6$GeV$^4$ if $\lambda= 100$GeV. However, the dynamics of our Universe suggest a vacuum energy density $\rho_\Lambda$ cannot be much greater than the critical energy $10^{-46}$GeV$^4$. This is either 117 or 52 orders of magnitude smaller than the theoretically natural one. This is the old cosmological constant problem:

(a) Why the cosmological constant is small?

Moreover, recent observations show there indeed exists a component of dark energy, with energy density indeed of order $10^{-46}$GeV$^4$. The observation made the old cosmological constant problem harder to solve. Because a small yet non-zero number is even more difficult to explain than an unexpected zero. Also, as dark energy has different dynamics (dilution rate from cosmic expansion) from matter or radiation, another problem emerges:

(b) Why dark energy density is the same order as matter density today?

The cosmological constant problems (a) (b) are so hard that some people gave up a fundamental explanation of the cosmological constant and treat it as an environmental variable; On the other hand, the problem is also so inspiring that numerous models have been built seeking for an explanation. In the remainder of this section, we shall overview eight classes of those attempts.

\subsection{Symmetry}
\label{sec:symmetry}

Symmetry has been promoted to a first principle in modern physics. It could not be better if the cosmological constant problem could be solved by a symmetry principle directly. SUSY \cite{wb} gets near in this aspect \cite{ZuminoBG}. But unfortunately SUSY has to be broken on large distance scales.

Nevertheless, there are a few proposals to apply symmetry principles to the cosmological constant problem. For example, 't Hooft and Nobbenhuis \cite{thooft} proposed that one could consider the following transformation for coordinate and momentum:
\begin{align}
  x^\mu \rightarrow iy^\mu, \quad p_x^\mu \rightarrow -ip_y^\mu.
\end{align}
A massless scalar field or an Abelian gauge field has this symmetry. On the other hand a cosmological constant is forbidden. However, the mass of a scalar is not allowed either (a positive mass squire transforms into a tachyon). Also non-Abliean gauge theory does not have such a symmetry. There could be issues about boundary conditions in quantum mechanics after the above transformation as well.

Instead of transforming coordinate, one could as well change the sign of the metric. In a theory with $g_{\mu\nu}\rightarrow -g_{\mu\nu}$ symmetry \cite{ErdemYD}, the cosmological constant is absent. Unfortunately it is hard to realize this symmetry in four dimension spacetime. As another attempt \cite{wett}, one could as well scale the metric instead of changing its sign. But there is an instability in such a theory.

As another class of approaches, Kaplan and Sundrum \cite{ks} considered a counter term for the cosmological constant as
\begin{align}
\label{ksl}
{\cal L}=\sqrt{-g}\left({M_p^2\over 2}R-\Lambda\right)
+{\cal L}_{\rm matter}(\psi,
D_\mu)-{\cal L}_{\rm matter}(\tilde{\psi},D_\mu),
\end{align}
where the Lagrangian for matter field $\psi$ and $\tilde\psi$ have the same function form. Again there is an issue of ghost instability considering the $\psi$ and $\tilde\psi$ fields are coupled by gravity.

\subsection{Anthropic principle}
\label{sec:anthropic-principle}

It is hopeless, if not impossible, to explain every parameter in nature from the first principles. For example, one cannot hope to
calculate the temperature on the earth from any theory of everything. There are a vast number of other planets in our
Universe, whose surface temperature is hugely different from that of the earth. And from the viewpoint of a theory of everything, the
earth is not too special from any other planets to have a special temperature.

However, a natural explanation of the earth temperature emerges, from the fact that there are intelligence livings on the earth, who could ask any question. This fact makes the earth special. Thus the temperature should fit for intelligent beings. This explanation is known as the anthropic principle \cite{carter} (of the weak sense \cite{DickeGZNature} due to Weinberg's classification \cite{weinrev}). Such an anthropic explanation of a parameter in nature makes sense if:

(1) The parameter could vary in different parts of the Universe. On the one hand, this makes room for an anthropic explanation. On the other hand, this implies the value of the parameter can not be calculated from the first principles (although its distribution may be calculated).

(2) The existence of intelligence is sensitive to this parameter, and the observed value (at the time of observation) prefers such an existence.

(3) Additionally, to relate an intelligent observer to us humans, one has to assume we are typical observers, and propose a measure to define the typicality in a probably infinite Universe.

The above are the general considerations of the anthropic principle. Those considerations may be applied to the cosmological constant problem. This is because:

(1) There might be different cosmological constants in different parts of the Universe, in space scopes beyond our observable Universe. This statement is (so far) not testable by the (current) definition of observable Universe. However, fundamental theories suggest such a possibility. Especially, in string theory, in the past decade a landscape of (semi-classical) meta-stable vacua solutions are discovered \cite{SusskindKW}\cite{bpkklt}. The number of such vacua in string theory may be of order $10^{500}$\cite{DouglasUM}, considering one could use different Calabi-Yau manifolds to compactify the extra dimensions, and add different branes and fluxes \cite{gkp}. Moreover, the theory of eternal inflation \cite{eternalSV}, as well as the quantum mechanical many world conjecture, open up possibilities to populate the string landscape. If the above picture holds, the cosmological constant indeed ends up as an environmental variable.

(2) The existence of intelligence is indeed sensitive to the cosmological constant. This is because a large and positive cosmological constant will dilute the structure of the Universe before useful structures (like galaxies) are formed \cite{anthrwein}. Whereas a large (in absolute value) and negative cosmological constant will drive an early collapse of the Universe. However, it is still too early to conclude that
our observed value at the time of observation is most suitable for intelligence. For example, there is a Boltzmann brain problem \cite{Boltzmann}, indicating that a (nearly) pure de Sitter Universe may be more suitable for intelligence, of a thermal fluctuation type.

(3) Finally, it is in debate if we can typically consider ourselves as typical observers, and consider the observed Universe as a typical one in the multiverse. Moreover, the measure problem \cite{measureproblem} is still severe. Currently several measures coincide with each other and a few other measures are ruled out. Thus it seems that different
efforts converge in the right direction. However, no guiding principle is available so far for the measure construction thus the problem should not be marked as solved.

\subsection{Tuning mechanisms}
\label{sec:tuning-mechanisms}

Another direction for dark energy is to introduce a field, which could dynamically cancel the cosmological constant, no matter how large it was originally. Such kind of a scenario is called tuning mechanism. For example, for the old cosmological constant problem, one may have expected to have a scalar field $\phi$, which is related to the stress tensor as \cite{DolgovGH}
\begin{align}
  \nabla^2\phi \propto T^\mu_\nu \propto R~.
\end{align}
If $T^\mu_\mu$ also depends on $\phi$ and drives $\phi$ to some stable value $\phi_0$, the equilibrium state at $\phi=\phi_0$ will have a vanishing Ricci scalar thus a vanishing cosmological constant.

Unfortunately, Weinberg \cite{weinrev} showed that it is difficult to realize such a mechanism. The above conditions are proven to be equivalent to the statement that the matter Lagrangian takes the form
\begin{align}
  \mathcal{L} = e^{4\phi} \sqrt{-g}  \mathcal{L}_0.
\end{align}
Thus unless one fine tunes $\mathcal{L}_0$, there will not be a stationary point for $\phi$. Also, if $\phi_0$ asymptotically rolls to $-\infty$, the effective Newton's constant $G_N$ vanishes, because $\phi_0$ can be viewed as a redefinition of $G_N$. Neither of the above possibilities is satisfactory.

There are some recent attempts trying to get around the above problem. For example, one can consider a non-trivial extra dimension and then Weinberg's proof no longer applies \cite{EDtunning}. However, attempts in this direction also encounter problems such as a singularity in the bulk, or violation of the Lorentz symmetry.

\subsection{Modified gravity}
\label{sec:modified-gravity}

Dark energy (as well as dark matter) is identified via gravitational probes.
Thus once one meets difficulties in the explanation of dark energy, it is a natural idea to step back and question if the current gravity theory can be applied to cosmological scales (or even galactic scales).

There have been a large number of theories of modified gravity, and a large proportion of which has been used as models of dark energy. An incomplete list of which include $f(R)$ gravity \cite{BuchdahlZZ}\cite{CarrollWY}\cite{fRReviews}, MOND \cite{MilgromCA} and TeVes \cite{BekensteinNE}, DGP \cite{DGPDGP}, scalar-tensor theories \cite{BransSX}\cite{AmendolaQQ}, Gauss-Bonnet \cite{ZwiebachUQ}\cite{NojiriVV} and Lovelock \cite{LovelockYV} gravities, Horava-Lifshitz gravity \cite{HoravaUW}, $f(T)$ gravity \cite{fTBL}, conformal gravity \cite{MannheimBFA}, fat graviton and so on. Those scenarios open up a huge variety of possibilities for dark energy.

The attempts of modifying gravity starts as early as the early 1960s, known as the Brans-Dicke gravity \cite{BransSX}. In Brans-Dicke gravity, a scalar field is added into the gravitational action
\begin{align}
  S = \int d^4 x \sqrt{-g}
  \left(
    \frac{1}{2}\phi R - \frac{\omega}{2\phi}(\partial\phi)^2
  \right)~.
\end{align}
Instead of inserting a scalar field, one could also modify the Einstein-Hilbert action by using an alternative curvature invariant. As an example, $f(R)$ gravity modifies the Einstein-Hilbert action as
\begin{align}
  \frac{M_p^2}{2} \int d^4 x \sqrt{-g} ~ R
  \quad\rightarrow\quad
  \frac{M_p^2}{2} \int d^4 x \sqrt{-g} ~ f(R)~.
\end{align}To make specific predictions one has to specify $f$. For instance, one could choose $f(R) = R - \mu^4/R$, where $\mu$ is a constant \cite{CarrollWY}.

Along this direction, one could also construct more general curvature invariants. For example, the Gauss-Bonnet term
\begin{align}
  R^2_\mathrm{GB} \equiv
  R^{\mu\nu\rho\sigma}R_{\mu\nu\rho\sigma}
  - 4 R^{\mu\nu}R_{\mu\nu} + R^2~.
\end{align}In 4 spacetime dimensions $R^2_\mathrm{GB}$ is a total derivative thus does not modify the local dynamics of gravity. However, one could couple $R^2_\mathrm{GB}$ to a scalar field as $f(\phi)R^2_\mathrm{GB}$ as a model of dark energy \cite{NojiriVV}.

The DGP scenario is another popular possibility of modifying gravity \cite{DGPDGP}. In the DGP scenario, two gravity theories are considered at the same time: one located on a 3+1 dimensional brane, and another on a 4+1 dimensional brane, with their own Einstein-Hilbert actions (while the two gravity theories share the same metric: the 3+1 metric is the 4+1 metric restricted on the 3-brane). The existence of the 4+1 brane leads to large scale modifications of gravity.

\subsection{Quantum cosmology}
\label{sec:quantum-cosmology}

The cosmological constant problem is a problem from the energy of quantum fluctuations, which gravitate. Thus it is not surprising if the solution could have something to do with quantum gravity, even though the problem mainly affects extremely large scales.

For example, Hawking \cite{hh}\cite{Hawking:1984hk} provided a mechanism to show that the cosmological constant is probably zero. He constructed a 4-form field strength, which could vary and change the cosmological constant. Then the wave function of the Universe indicates a probability distribution of the effective cosmological constant $\Lambda$ (before normalization of the probability distribution) to be
\begin{align}
  P(\Lambda) = e^{\frac{3\pi}{G\Lambda}}.
\end{align}
Thus $\Lambda\rightarrow 0$ from above is the most probable value of the cosmological constant when the Universe is created via the Hartle-Hawking wave function. Unfortunately, this scenario produces an empty Universe, with only a vanishing cosmological constant. On the one hand, there is no radiation or matter in this approach, and on the other hand, our present accelerating Universe (if not coming from another source) falsifies the mechanism.

Later, the probability is corrected by considering the decoherence effects \cite{FiSa}. In this case the Universe could be born at the string scale. However, in this case this mechanism is no longer very related to the cosmological constant problem. Instead, the mechanism is used as a program to select the original Universe.

As another quantum approach to dark energy, Tye \cite{HenryTyes} considered a possibility that the vacuum of our Universe may experience fast tunnelings through the string landscape, until settling down at a place with a small cosmological constant. The idea is supported by the resonant tunneling effect in quantum mechanics. Resonant tunneling is a well established effect in quantum mechanics. Recent development also shows evidence that resonant tunneling can exist in quantum field theory \cite{CopelandQF}\cite{SaffinVI}\cite{Tye:2009rb}\cite{Tye:2011xp}.

One could also relate quantum features of cosmology to dark energy from a cosmological seesaw mechanism \cite{MotlsSeesaw, CarrollSeesaw}. It is interesting to notice that if SUSY is broken at the TeV scale, there is a hierarchy of scales
\begin{align}
\label{QChierarchy}
M_{\Lambda} \sim M_{\rm SUSY}^2
/ M_{p},
\end{align}
And the SUSY scale could also be replaced by any scale that stabilizes Higgs mass. It is argued that the hierarchy could be resolved by a seesaw mechanism relating two interacting Universes. The coupling is introduced through the wave function of the Universe \cite{McGuiganHS}. It is so far not clear yet how to introduce such a coupling at a more fundamental level.

\subsection{Holographic principle}
\label{sec:hologr-princ}

Effective field theory is one of the most important concepts in high energy physics. The cosmological constant problem, as a problem of renormalizing the cosmological constant, can also be thought of as a problem in the sense of effective field theory. However, the application regime for effective field theory seems not as well understood as people thought years ago: Not only an increase of energy scale, but also an increase of the number of relevant degrees of freedom, seems to lead to a break down of effective field theory.

There are quite a few examples for the above observation. Here we only consider the most relevant one for our purpose: Consider the vacuum energy within a volume with radius $L$ \cite{CohenZX}. The volume under consideration should not form a black hole thus the effective field theory should have a cutoff $\Lambda_\mathrm{UV}$ satisfying
\begin{align}
  L^3 \Lambda_{UV}^4 \sim E \leq L M_p^2~,
\end{align}
where the very right hand side is the energy of a black hole of the same size. This inequality can be recast in terms of energy density as
\begin{align}\label{eq:hde-density}
  \rho = 3c^2 M_p^2 L^{-2}~,
\end{align}
where $c$ is a parameter introduced in \cite{LiRB}, which is naturally a constant of order one.

As pointed out in \cite{LiRB}, when $L$ is chosen the event horizon of our Universe, $\rho$ behaves as a component of dark energy, dubbed the holographic dark energy (HDE). The equation of state for this component is
\begin{align}
  w_\mathrm{hde} = -\frac{1}{3}-\frac{2\sqrt{\Omega_\mathrm{hde}}}{3c}~,
\end{align}
where $\Omega_\mathrm{hde}$ is the relative energy density of HDE. Thus when HDE subdominates, it behaves as curvature; and when it dominates the Universe, the Universe behaves as quasi-de Sitter for reasonable $c$.

The equation (\ref{eq:hde-density}) can be also understood in several other ways. For example, as shown in \cite{LiPMZY}, the Casimir energy in de Sitter space has a quadratic divergence. Thus the Casimir energy takes the form (\ref{eq:hde-density}) and de Sitter expansion may be self-driven. Alternatively, if one considers the event horizon of our Universe as a holographic screen, the acceleration of Universe behaves as an entropic force \cite{LiCJ}, or a consequence of quantum information \cite{LeeZQ}. Also, if non-perturbative gravitational excitations behaves as a gas component with holographic dispersion relation, one could also arrive at (\ref{eq:hde-density}) \cite{LiQH}.

The event horizon is a natural choice, but not the unique logical possibility. It is also proposed that the Ricci scalar \cite{RDEPaper}, or the time of the Universe \cite{CaiUSWeiTY}, could be used as the IR cutoff as well. Also the parameter $c$ could be a function of time or redshift \cite{Zhang:2012qr}.

\subsection{Back-reaction}
\label{sec:back-reaction}
The Einstein equations are non-linear. The non-linearity brings a great amount of difficulties as well as challenges to applications of the theory of gravity.

One of the distinguishing features of a non-linear theory is that the time evolution and the operation of averaging may not commute. In the conventional approach of cosmology, we first approximate our Universe to be homogeneous and isotropic. This is true only on large scales. We are averaging over smaller structures before considering time evolution of the scale factor. Thus there will be a correction from the non-commutivity of averaging and time evolution. The correction is dubbed back-reaction \cite{RasanenKI}, and the question is if the correction could become large and accelerate our Universe.

The back-reaction term appears in the cosmological equations as \cite{RaychaudhuriYV}
\begin{align}
\label{BRavgeRay}
3{\ddot {a} \over {a}} = -4\pi
G\langle\rho\rangle+{\cal Q},
\end{align}
\begin{align}
\label{BRavgeFried}
3 H^2 = 8\pi
G\langle\rho\rangle-{1\over 2}\langle R_3\rangle-{1\over 2}{\cal Q},
\end{align}
\begin{align}
\label{BRavgeCont}
\partial_t\langle\rho\rangle+3H\langle\rho\rangle=0,
\end{align}
where $R_3$ is the spatial curvature, and $\cal Q$ is defined as a combination of the variances of the expansion scalar $\theta$ and the shear $\sigma$ as
\begin{align}
\label{BRQDef}
{\cal Q}
\equiv {2\over
3}\left(\langle\theta^2\rangle-\langle\theta\rangle^2\right)
-2\langle\sigma^2\rangle.
\end{align}
It is still an open question if $\cal Q$ could be large enough to drive the cosmic expansion. Also, in this approach, one still has to solve the old cosmological constant problem, to have a not-too-large original cosmological constant.

One could also consider whether super-Hubble inhomogeneities could back-react the evolution of our Universe.
This is a more ambitious approach.
Because if there is a screening effect of cosmological constant form the back-reaction,
one can solve the old cosmological constant problem together.
One can prove \cite{UnruhIC}\cite{GeshnizjaniWP} that if there is only curvature fluctuation,
the super-Hubble inhomogeneities cannot back-react onto our Hubble scale dynamics.
When there is isocurvature fluctuation, the effect of super-Hubble back-reaction is still in debate \cite{SHBRdebate}.

\subsection{Phenomenological models}
\label{sec:phen-models}
As another huge class of attempts, dark energy could as well be a simple fluid component, for example, a scalar field. The simplest example of this class is a quintessence field with a conventional scalar \cite{WetterichFM,ZlatevTR}
\begin{align}
\label{PMquintessenceAction}
S = \int
d^4 x \sqrt{-g} \left[-{1\over
2}g^{\mu\nu}\partial_\mu\varphi\partial_\nu\varphi
-V(\varphi)\right],
\end{align}
At the homogeneous and isotropic background level, the energy density and the pressure takes the form
\begin{align}
\label{PMquintessenceRhoP}
\rho = {1\over
2}\dot\varphi^2+V(\varphi), \qquad p = {1\over
2}\dot\varphi^2-V(\varphi).
\end{align}
When the potential $V$ is flatter than
\begin{align}
\label{PMcriticalPotentialAGAIN}V=V_0 \exp \left(
-{\sqrt{2}\varphi\over M_p}\right)
\end{align}
an accelerating solution for our Universe is admitted. In the limit when the potential $V$ is completely flat, the scenario returns to a massless scalar field, plus a decoupled non-dynamical cosmological constant. A quintessence field always has equation of state $w>-1$.

The quintessence scenario could be modified in various ways. For example, one could consider a scenario of phantom, which introduces a quintessence field with a wrong-sign kinetic term \cite{CaldwellEW}. In this case, one get a field with $w<-1$ as dark energy. The phantom field is unstable when coupled to matter (even gravitationally). Even worse, in a Lorentz invariant theory, the decay rate of phantom is divergent \cite{Cline:2003gs}. Recently, it is argued that a dynamical breaking of Lorentz invariance due to the decay of phantom itself may avoid this problem \cite{Garriga:2012pk}. The combination of quintessence and phantom fields also makes a scenario of quintom \cite{Quintom}.

As another direction, one could generalize the standard kinetic term as \cite{kessence}
\begin{align}
  -{1\over
2}g^{\mu\nu}\partial_\mu\varphi\partial_\nu\varphi
\rightarrow
P\left( -{1\over
2}g^{\mu\nu}\partial_\mu\varphi\partial_\nu\varphi , \varphi\right),
\end{align}
where $P$ is a general function of its two arguments. This generalization is known as k-essence. The scenario can be further generalized to ghost condensation \cite{ArkaniHamedUY, PiazzaDF} and kinetic gravity braiding \cite{KGB}.

The field theoretic approach of dark energy can also be generalized to spinor fields \cite{spinorDE}, vector fields \cite{vectorDE}, and $p$-form fields \cite{pformDE}. Also, dark energy could be particles with exotic properties \cite{LiQH}\cite{particleDE}.

As another class of phenomenological models, dark energy may as well be a fluid with a modified equation of state. For example, Chaplygin gas \cite{AbramoDB,KamenshchikCP} is a model in this direction. Chaplygin gas, originally introduced to model the air flow around the wing of an aircraft, is a fluid with equation of state
\begin{align}
  p=-A/\rho~.
\end{align}
Later, the scenario is generalized to \cite{GCGas}
\begin{align}
  p=-A/\rho^\alpha
\end{align}
to better fit the data and to unify dark matter and dark energy.

As another example of fluid dark energy, dark energy might be a fluid with viscosity \cite{BrevikSJBJ}
\begin{align}
  T_{\mu\nu}=\rho u_\mu u_\nu + (p-3H\zeta) (g_{\mu\nu}+u_\mu u_\nu)~.
\end{align}

There are two generic features for most phenomenological models of dark energy. Theoretically, one still need to worry about the cosmological constant problem because the large vacuum zero point energy is not canceled in a natural way. Phenomenologically, one should consider dark energy perturbations as well in those models, because dark energy has local degrees of freedoms in those cases. The latter could be a window to distinguish field, fluid, particle scenarios of dark energy between a simple cosmological constant.

\section{Observational aspects}

\subsection{Cosmic probes of Dark Energy}

The most common approach to probe dark energy is through its effect on the expansion history of the Universe.
This effect can be detected via the luminosity distance $d_L(z)$ and the angular diameter distance $D_A(z)$.
In addition, the growth of large-scale structure can also provide useful constraints on dark energy.
In this section, we briefly review some mainstream cosmological probes of dark energy.
For a more detail review on current mainstream observational probes, see Ref. \cite{ReviewofProbes}.

\subsubsection{Type Ia supernoave}

A white dwarf star can accrete mass from its companion star; as it approaches the Chandrasekhar mass,
the thermonuclear explosion will occur, representing itself as a type Ia supernoave (SNIa).
This mechanism implies that SNIa can be used as standard candles to measure the luminosity distance $d_L(z)$,
which takes the form $d_L(z)=\int^z_0\frac{dz^\prime}{H(z^\prime)}$.
Clearly, $d_L(z)$ contains the information of the cosmic expansion history $H(z)$ through an integration.

In 1998, using the SNIa data from the Hubble space telescope (HST) observations,
Riess {\it et al.} \cite{riess} first discovered that the expansion rate of our Universe is accelerating.
Soon after, the result was confirmed by Perlmutter {\it et al.} \cite{perl} based on the analysis of 42 SNIa.
The discovery of the Universe accelerating expansion was named ``Breakthrough of the Year'' by Science Magazine in 1998.
After that, the surveys of SNIa has drawn lots of attention.
Famous teams focusing on this field include the High-$z$ team \cite{Goldfour},
the Supernova Legacy Survey (SNLS) \cite{SNLS}, the ESSENCE \cite{ESSENCE},
the Sloan Digital Sky Survey (SDSS) Supernova (SN) Survey \cite{SDSSN}, and so on.
Recent SNIa dataset includes the three-year SNLS dataset \cite{SNLSConley} consisting of 472 SN
and the Supernovae Cosmology Project (SCP)'s Union 2.1 dataset \cite{UnionTwo} consisting of 580 SN.
When combined with other cosmological probes,
these data could reduce the (1$\sigma$) error of the dark energy EoS to less than 0.1.

For a SNIa, once its redshift $z_i$, absolute magnitude $M_i$ and apparent magnitude $m_i$ are determined,
one can construct the theoretical (assuming specific dark energy) and observational distance modulus,
\be
\mu_{obs,i}=m_i-M_i,\ \ \ \mu_{th,i}=5\log _{10}d_L(z_i)+25,
\ee
and the $\chi^2$ for the SNIa data can be constructed as
$\chi^2_{SN}=\sum_i \frac{[\mu_{obs,i}-\mu_{th,i}]^2}{\sigma^2_i}.$
In this procedure, people often analytically marginalize the nuisance Hubble constant $H_0$ \cite{SNNuisance}.

Due to the absence of the knowledge of its detailed explosion mechanism, SNIas are not intrinsically standard candles.
Other systematic errors of the SNIa come from the photometry, the identification of SNIa,
the host-galaxy extinction, the gravitational lensing, and so on.
These systematic errors are now the major factor that confines the precision of SNIa data.
To enhance the precision,
improvements on the photometric technique, as well as better understandings of the dust absorption and the SN explosions, are needed.

\subsubsection{Cosmic Microwave Background}

As the legacy of the cosmic recombination epoch, the cosmic microwave background (CMB) contains abundant information of the early Universe.
Current CMB experiments include the well-known Wilkinson Microwave Anisotropy Probe (WMAP) \cite{WMAP},
the Planck satellite \cite{DEObjPLANCK}, and so on.

From the CMB observations, some distance priors can be extracted to constrain dark energy.
For example, the ``shift parameter'' \cite{shiftparameter} defined at the decoupling epoch $z_\ast$ takes the form
$R(z_{\ast}) = \Omega_m H_0 (1 + z_{\ast} )D_A (z_\ast)$.
This parameter nicely describes the full CMB leverage on the cosmic expansion history,
and is a good approximation for models not too much deviated from $\Lambda$CDM.
Another distance ratio is the ``acoustic scale'' $l_A$, defined as
$(1+z_{\ast})\frac{\pi D_A(z_\ast)}{r_s(z_\ast)}$
($r_s$ is the sound horizon).
This quantity represents the CMB multipole of the acoustic peak location.
As high redshift distance indicators,
$R$ and $l_A$ are especially useful in breaking the degeneracies between cosmological parameters when combined with other cosmological probes.

The CMB data could also be used to probe dark energy through the Integrated Sachs Wolfe (ISW) effect \cite{CMBISW}.
This large scale anisotropy effect is caused by the variation of the gravitational potential during the epoch of the cosmic acceleration,
and has been detected through the cross relation between CMB and the LSS at about 4$\sigma$ \cite{CMBShirley}.

\subsubsection{Baryon Acoustic Oscillations}

Baryon acoustic oscillations (BAO) refers to overdensities or clusterings of baryonic matter
at certain length scales ($\sim$150 Mpc in today's Universe) due to acoustic waves which propagated in the early Universe
\cite{BAOPeeblesandYu, BAOSilk}.
It provides a ``standard ruler'' for cosmological observations,
and can be measured at low redshifts $z<1$ through galaxy surveys.
Famous BAO measurements include the Two-degree-Field Galaxy Redshift Survey (2dFGRS) \cite{BAOColless},
the SDSS \cite{BAOYorkDG}, the WiggleZ dark energy survey \cite{BAOWiggleZ}, and so on.
When combined with the SNIa and CMB data,
the latest BAO data from the WiggleZ survey yields a result $w=-1.03\pm0.08$ \cite{BAOWiggleZData}.

From galaxy observations, the BAO scales in both transverse and line-of-sight directions are obtained,
corresponding to the quantities $r(z)/r_s(z_d)$ ($r(z)$ is the comoving distance; $z_d$ is the ``drag epoch'') and $r_s(z_d)/H(z)$.
Widely used BAO distance measurements include the $A$ parameter, the distance $D_V(z)$, the distance ratio $r_s(z_d)/D_V(z)$, and so on.
Current measurements distribute at the redshift region $z\sim0.1-0.75$.

Since the BAO measurement only requires the determination of the three dimensional positions of galaxies,
they are less affected by astronomical uncertainties.

\subsubsection{Weak lensing}

Weak lensing (WL) is the slight distortions of distant objects' images,
due to the gravitational bending of light by structures in the Universe.
The mass and positions of the lens depend on the matter distribution on the light cone,
while the distances to the objects and lens are determined by the geometry of the spacetime.
So WL probes dark energy through its effects on both the cosmic expansion and growth history of the strucuture.
The current WL project is the Canada-France-Hawaii legacy survey (CFHTLS) \cite{WLCFHTLS}.

The effect of WL on the distance objects represent on the distortions of the shapes of the galaxies \cite{WLReview}
\foot{It should be noted that the lensing of CMB can also be used to probe dark energy.
In 2011, using the Atacama Cosmology Telescope lensing measurements,
Sherwin {\it et al.} \cite{WLACT} reported that the CMB measurement alone favors cosmologies with $w=-1$ dark energy
over models without dark energy at a 3.2$\sigma$ level.},
with typical scale $\sim$0.01.
This signal is much smaller than the typical scale of the intrinsic deviation in the galaxy shape (about 0.3-0.4) ,
and can be only detected from a large number of galaxies.
The most commonly used method is the cosmic shear statistics.
In 2005, based on the cosmic shear data of CFHTLS, Hoekstra {\it et al.} \cite{WLHoekstra} reported a constraint of $w<-0.8$ at the 68\% CL.

Although current WL data are not powerful enough compared with SNIa and BAO,
WL is a very promising probe to detect dark energy.
The underlying WL data are rich,
while the observational and statistical techniques of the WL probe are under continually improvement.

\subsubsection{Other dark energy probes}

 {\it Galaxy clusters:} Their distribution marks the LSS.
Since the comoving volume of the Universe is affected by the geometry,
the cluster abundance also encodes the information of the expansion history \cite{CLHaiman}.

 {\it Gamma-ray bursts:} As the most luminous electromagnetic events in the Universe,
gamma-ray bursts (GRBs) may help us to extend the Hubble diagram to $z\sim8$ \cite{GRBSchaefer}.
However, due to the large errors and calibration problem,
current GRB data are not able to provide extensive and reliable probe to dark energy.

{\it X-ray observations:} Like the SNIa and BAO data, the X-ray $f_{gas}$ (defined as $M_{gas}/M_{tot}$) measurements
can also probe the redshift-distance relation,
with the dependence $f_{gas} \propto d_L(z)D_A(z)^{0.5}$.
In 2008, based on the Chandra X-ray data,
Allen {\it et al.} \cite{XRayAllenZeroEight} reported $\Omega_{\Lambda}=0.86\pm0.21$.

 {\it Hubble parameter measurements:} Precision measurements of $H_0$ can help to break the degeneracies,
and are thus widely used in dark energy constraints.
In \cite{HzRiessnew}, Riess {\it et al.} reported $H_0 = 73.8 \pm 2.4 {\rm km/s/Mpc}$.
Besides, the direct measurement of $H(z)$ from the differential ages of passively evolving galaxies
can also provide valuable constraints \cite{HzSimon}.

 {\it Cosmic age test:} Cosmic age problem is a ``smoking-gun'' of evidence for the existence of dark energy
\foot{In the matter-dominated SCDM model, the present age of the Universe is only $t_0=\frac{2}{3H_0}=9$Gyr, and
contradicts with the detections of many older than 10Gyr objects in astronomical observations.}.
However, the existence of some extremely old globular clusters and quasars shows that
the cosmic age problem has not been completely removed by the introduction of dark energy \cite{HzMys,AgeSW}.
In addition, the ``look back time-redshift'' \cite{HzSimon} data obtained from the age of old passive galaxies
can be included into the $\chi^2$ analysis and constrain dark energy.

{\it Growth factor data:} The growth factor defined by $f=d\ln\delta/d\ln a$ can be measured from the redshift distortions and Lyman-$\alpha$ Forest data.
Current growth factor data \cite{GrowthFacData} have large errors and are still less powerful.

\subsection{Dark energy projects}
\label{secDEProjects}

According to the DETF report \cite{DETF}, the dark energy projects can be classified into four stages:
completed projects are Stage I;
on-going projects, either taking data or soon to be taking data, are Stage II;
intermediate-scale, near-future projects belong to Stage III;
larger-scale, longer-term future projects belong to Stage IV.
In this paper, we will introduce some mainstream Stage IV projects.
For a comprehensive list of the dark energy experiments, see Ref. \cite{ReviewofTurner,ReviewofOurs}.

\subsubsection{Large Synoptic Survey Telescope}

A most ambitious ground-based dark energy survey project is the Large Synoptic Survey Telescope LSST \cite{DEObjLSST},
which is an USA NSF/DOE collaboration program.
Its design is an 8.4-meter ground-based optical telescope to be sited in Cerro Pachon of Chile.
Over a 10-year lifetime, LSST will obtain a database including 10 billion galaxies and a similar number of stars,
studying dark energy through a combination of the BAO, SN and WL techniques.
Since its compelling scientific capacity and relatively low technical risk,
LSST was selected as the top priority large-scale ground-based project for the next decade of astronomy in the Astro2010 report \cite{AstroLatest}.
The project is scheduled to have first light in 2016 and the beginning of survey operations in 2018.

\subsubsection{Square Kilometer Array}

Another ambitious large-scale ground-based telescope is the Square Kilometer Array (SKA) \cite{DEObjSKA},
an international collaboration program.
SKA has a collecting area of order one square kilometer and is capable to operate at a wide frequency range
(60 MHz - 35 GHz).
It will be built in the southern hemisphere (either in Australia or in South Africa),
and the specific site will be determined in 2012.
SKA will probe dark energy by BAO and WL techniques via the measurements of the Hydrogen line (HI) 21-cm emission in normal galaxies at high redshift.
Its construction is scheduled to begin in 2016 for initial observations by 2020 and full operation by 2024.

\subsubsection{Wide Field Infrared Survey Telescope}

A most exciting space-based dark energy project is the Wide Field Infrared Survey Telescope (WFIRST) \cite{DEObjWFIRST},
which is an USA NASA/DOE collaboration program.
WFIRST is a 1.5-meter wide-field near-infrared space telescope,
orders of magnitude more sensitive than any previous project.
It will determine the nature of dark energy by measuring the expansion history and the growth rate of large scale structure.
A combination of the BAO, SN and WL techniques will be used.
Since its compelling scientific capacity and relatively low technical risk,
WFIRST was selected as the top priority large-scale ground-based project for the next decade of astronomy in the Astro2010 report \cite{AstroLatest}.
The project is scheduled to launch in 2020 and has a 10-years lifetime.

\subsubsection{Euclid}

Another exciting space-based dark energy project is the Euclid \cite{DEObjEuclid},
which is an ESA project.
Euclid is 1.2 m Korsch telescope, with optical and near-infrared (NIR) observational branch.
It will also make use of several secondary cosmological probes
such as the ISW, CL and redshift space distortions to provide additional measurements of the cosmic geometry and structure growth.
After 5 years' survey, Euclid will measure the DE EoS parameters $w_0$ and $w_a$ to a precision of $2\%$ and $10\%$, respectively,
and will measure the growth factor exponent $\gamma$ with a precision of $2\%$.

\subsubsection{International X-ray Observatory}

The last Stage IV project is and the International X-ray Observatory (IXO) \cite{DEObjIXO},
which is a partnership among the NASA, ESA, and the Japan Aerospace Exploration Agency (JAXA).
It is a powerful X-ray space telescope that features a single large X-ray mirror assembly
and an extendable optical bench with a focal length of $\sim20$ m.
With more than an order of magnitude improvement in capabilities,
IXO will study dark energy through the WL technique.
It was selected as the fourth-priority large-scale ground-based project in the Astro2010 report
\cite{AstroLatest},
and is scheduled to launch in 2021.

\subsection{Observational Constraints on Theoretical Models}

In this section we briefly review some research works concerning the cosmological constraints on various theoretical models.

\subsubsection{Scalar field models}

As the most popular phenomenological dark energy models,
scalar filed models can mimic a dark energy component with arbitrary properties (e.g., arbitrary $w$).
So the issue of the observational constraints on the scalar field models is somewhat similar to the observational probe of the dynamical behavior of dark energy.

Different classes of scalar field dark energy models have different properties.
For  quintessence or phantom,  $w>-1$ or $w<-1$,
while for quintom $w$ can cross $-1$
\foot{The value of dark energy EoS $w$ is crucial to the future fate of our Universe.
The Universe will end up with a ``big rip'' if $w<-1$ \cite{BigRip}.}.
Current observations still can not determine whether dark energy is quintessence, phantom or quintom.
For example, the Seven-year WMAP measurements \cite{WMAPSeven} gave $w=-1.10\pm0.14$, with all the possibilities exist.

Another popular issue often discussed in the literature of the scalar field models is possible interactions between dark energy and dark matter \cite{DEIntWetterich}.
Phenomenologically, interaction can be introduced in the form
\be
\dot\rho_m+3H\rho_m=Q,\ \ \ \dot\rho_{de}+3H(\rho_{de}+p_{de})=-Q,
\ee
where $Q$ is the interaction term.
Current data put tight constraints on interactions between the dark sectors.
In \cite{CQHe}, based on the analysis of the SNIa+BAO+CMB data,
Het {\it et al.} considered the form $Q=3\zeta H\rho$,
and obtained $\zeta\lesssim 0.01$ for the cases of $\rho=\rho_m$ and $\rho=\rho_m+\rho_{de}$.
Similar results were obtained in e.g. \cite{CQOther}.

\subsubsection{Chaplygin gas models}

The standard Chaplygin gas model with $p = A/\rho$ has been ruled out by the SNIa, BAO and CMB observations at more than 99\% confidence level (CL) \cite{NCGasNumWorks}\cite{DEManyModelsTDavis}.

The generalized Chaplygin gas model with $p = A/\rho^\alpha$ is once believed as a promising candidate to unify dark matter and dark energy (hereafter noted as ``UDM'').
The evolution of its energy density takes the form,
\be
\rho(t)=\rho_0[(1-\Omega_\star)+\Omega_\star a^{-3(1+\alpha)}]^{1\over(1+\alpha)},
\ee
where $\Omega_\star$ is the effective matter ratio determined by $A$ and and integration constant.
The model can nicely fit the observations of the cosmic expansion history when $\alpha\sim0$.
Especially, the case of the $\Lambda$CDM model corresponds to $\alpha=0$.

However, it is later proved that \cite{NCGasSandvik} this UDM scenario could produce oscillations or exponential blowup of the matter power spectrum,
which is inconsistent with observations.
In this way, the parameter space is confined to $|\alpha|\lesssim10^{-5}$ \cite{NCGasPark}, extremely close to the $\Lambda$CDM case.
So the generalized Chaplygin gas model as a UDM has been effectively ruled out.

\subsubsection{Holographic dark energy models}

In \cite{LiRB}, Li proposed the HDE model with the future event horizon as the IR cutoff.
Hereafter, a lot of works have been performed testing this models using the observational data of SNIa, BAO, CMB, WL, and so on.
These works all show that the HDE model can provide a good fit to the data.
For example, in \cite{HDECompareHDEModels}, by using the Consitution+BAO+CMB data,
Li {\it et al.} found that $\chi^2_{\Lambda {\rm CDM}}=467.775$ and $\chi^2_{\rm HDE}=465.912$,
implying that HDE is a competitive dark energy candidate \cite{HDEInteraction}.

Numerical studies of the HDE model show that the data favor $c<1$ \cite{HDEHuangSNIa, DEManyModelsMLi, HDEMaGongChen,DEManyModelsHWei}.
Recent studies have obtained the constraints $c<0.9$ at the 95\% CL.
The data also put tight constraints on possible interactions between HDE and dark matter.
Generally speaking, for an interaction term with the form $Q=3\zeta H\rho$ (where $\rho$ can be $\rho_m$, $\rho_{de}$, $\rho_{m}+\rho_{de}$, and so on), a constraint of $|\zeta|\lesssim0.01$ can be obtained based on the analysis of the SNIa+BAO+CMB data \cite{WangJX}.

As mentioned above, the Ricci scalar \cite{RDEPaper} and the time of the Universe \cite{CaiUSWeiTY} have also been proposed as the IR cutoff.
However, later researches showed that these two models are not favored by the data.
Compared with the original HDE model, the $\chi^2_{min}$ of these models can be 15-30 larger \cite{DEManyModelsHWei,HDECompareHDEModels,DEManyModelsMLi} based on the analysis of the SNIa+BAO+CMB data.

\subsubsection{DGP model}

Although it is a simple mechanism for the cosmic acceleration without introducing the dark energy component,
the DGP model can not provide a satisfactory fit to the data.
Firstly of all, the expansion history predicted by the DGP model leads to apparent inconsistency between the SNIa, BAO and CMB data  \cite{DEManyModelsTDavis},
implying that this model is not acceptable.
In \cite{DEManyModelsDRubin}, by using the Union+BAO+CMB data Rubin {\it et al.} showed that the DGP model leads to a $\Delta \chi^2_{min}=15$ compared with the $\Lambda$CDM model.
Secondly, the history of the structure formation in the DGP model is also disfavored by observations \cite{DGPLSTests}.
In \cite{DGPHorScalGrowGeo}, Fang {\it et al.} showd that this model over predicts the low-$l$ anisotropies of the CMB temperature power spectrum.
Compared with the $\Lambda$CDM, DGP is excluded at 4.9$\sigma$ (with curvature) and 5.8$\sigma$ (without curvature) levels.
Thus, the situation of the DGP model is far from optimistic when confronted with cosmological observations \cite{DGPDisfavored}.

\subsubsection{Other modified gravity models}

$f(R)$ models can mimic indistinguishable expansion histories as the $\Lambda$CDM model,
so to test them we need to lend help from the cosmic growth history observations \cite{fRTest}.
In \cite{fRSong}, Song {\it et al.} parametrized the $f(R)$ model by the ``Compton wavelength parameter'' $B$
\foot{The definition is $B=\frac{f_{,RR}}{f_{,R}}\frac{HdR/d\ln a}{dH/d\ln a}$. $B>0$ is the stable branch.},
and found that the model predicts a lower large-scale CMB anisotropies by reducing the ISW effect.
So far, tightest constraints on the $f(R)$ models come from the observations of the galaxy clusters.
In \cite{fRLombriser}, Lombriser {\it et al.} revisited the model studied in and reported $B_0<1.1\times 10^{-3}$ at the 2$\sigma$ CL,
mainly due to the inclusion of the cluster abundance data.

Gauss-Bonnet gravity predicts strong growth of perturbations on smaller scales,
incompatible with the observed galaxy spectrum,
unless the deviation from the Einstein gravity is very small \cite{GBKovisto}.
Thus this model has been effectively ruled out.

Based on the constraints on the PPN parameter $\gamma$ from the solar system and binary pulsar observations \cite{TESTGRAVCassini}\cite{TESTGRAVConfGRExp},
the Brans-Dicke theory has been tightly constrained to the level that its cosmological effects are insignificant.

\subsubsection{LTB models}

LTB models are the most commonly considered inhomogeneous scenarios to explain the cosmic acceleration without introducing dark energy.
In the simplest class of such models, we live close to the center of a spherically symmetric Gpc scale void,
and our local region has a larger Hubble parameter than the outer region due to the spatial gradients of the metric.
These models can successfully explain the dimming of the distant supernovae,
providing a nice fit to the SNIa data \cite{LTBBellidoB}\cite{DEManyModelsSollerman}, comparable with the $\Lambda$CDM model.

In \cite{LTBCenterA}, Alnes {\it et al.} pointed out that,
if it happened that we do not live precisely near (less than $\sim$20 Mpc \cite{LTBMoss}) the center of the void,
the observed CMB dipole would become much larger than that allowed by observations.
So there is a fine-tuning problem in the void models.
What is worse is that, even if we happen to live very close to the void center,
the existence of the off-center galaxy clusters will lead to an observationally kinetic Sunyaev-Zeldovich effect
due to their relative motion between the CMB frame.
In \cite{LTBkSZ}, Bellido {\it et al.} demonstrated that the limited observations of only 9 clusters with large error bars
already rule out LTB models with void sizes greater than 1.5 Gpc and a significant underdensity.

There have been many other interesting methods proposed to test the LTB models,
including the Hubble constant measurement \cite{HzRiessnew}, the redshift drift \cite{LTBTimeDrift},
the ionized Universe mirror \cite{LTBMirror}, the cosmic age test \cite{CosmicAgeLTB},
the constant curvature condition \cite{LTBClarksonA}, and so on.
The test of the LTB models is related to the test of the Copernican principle,
and this issue has drawn a lot of interests in recent years.

\subsubsection{Comparison of dark energy models}

The $\chi^2$ analysis alone cannot provide an effective way to make a comparison among dark energy models.
To enforce a comparison, a general way is to employ the information criteria (IC) \cite{ModelCompIC}.
The most frequently used IC are the Bayesian information criterion (BIC) \cite{ModelCompBIC}
and the Akaike information criterion (AIC) \cite{ModelCompAIC},
defined as,
\be
{\rm BIC}=-2\ln {\mathcal L}_{max}+k\ln N,\ \ \ {\rm AIC}=-2\ln {\mathcal L}_{max}+2k,
\ee
where ${\mathcal L}_{max}$ is the maximum likelihood (satisfying $-2\ln {\mathcal L}_{max}=\chi^2_{min}$ under the Gaussian assumption),
$k$ is the number of the parameters of the considered model, and $N$ is the number of data points used in the fit.
Clearly, these statistics favor models that give a good fit with fewer parameters.
Generally speaking, a $\Delta$BIC of more than 2 (or 6) could be considered as positive (or strong) evidence against a model.

In \cite{DEManyModelsTDavis}, based on the joint analysis of the SNIa+BAO+CMB data,
Davis {\it et al.} systematically studied a number of dark energy models and made a comparison among them.
Utilizing their AIC and BIC values, the ``rank'' of these models are determined.
The $\Lambda$CDM model achieves the best of all the models due to its nice fit to the data and the economy of parameters.
A series of models, including the constant $w$ model, the CPL parametrization and the Cardessian expansion,
can also provide comparably good fits but have more free parameters, thus less favored by the ICs.
Moreover, the dynamical behavior of these models all tend to collapse to the $\Lambda$CDM.
The DGP and the standard Chaplygin gas models are clearly disfavored.
More studies on the comparison among dark energy models have been performed in
with similar results obtained \cite{DEManyModelsMLi, DEManyModelsHWei, DEManyModelsDRubin, DEManyModelsSollerman, DEManyModels}.

In all, although some dark energy models have been ruled out (or shown to be less optimistic) based on the current cosmological observations,
we still need more observational data with higher precision to distinguish the various dark energy models.

\subsection{Model-independent dark energy reconstructions}
\label{secReconDE}

In addition to the cosmological constraints on specific dark energy models,
another route of numerical studies, i.e. the model-independent dark energy reconstructions, has also drawn more and more attentions.

The dark energy reconstruction is a classic statistical inverse problem for the Hubble parameter
\begin{align}
\label{HBasic}
H(z) = H_{0} \sqrt{\Omega_{m0}(1+z)^{3}+(1-\Omega_{m0})f(z)},
\end{align}
where $f(z)\equiv \rho_{de}(z) / \rho_{de}(0)$ is the dark energy density function.\index{dark energy density function}
Different reconstruction method will give different $f(z)$.
The main target of a dark energy reconstruction is to detect the dynamical property of dark energy.
It is widely believed that the EOS of dark energy $w_{de} \equiv p_{de}/\rho_{de}$ holds essential clues for the nature of DE,
because it is related with $f(z)$ through an integration
\begin{align}
\label{fw}
f(z) = \exp\left(3\int_{0}^{z}dz'{1+w_{de}(z')\over 1+z'}\right).
\end{align}

The model-independent dark energy reconstructions can be divided into four classes:
(i) Specific Ansatz: assuming a specific parameterized form for $w_{de}(z)$ and estimating the associated parameters.
(ii) Binned Parametrization: dividing the redshift range into different bins and using a simple local basis representation for $w_{de}(z)$ or $\rho_{de}(z)$.
(iii) Polynomial Fitting: treating the dark energy density function $f(z) \equiv \rho_{de}(z)/\rho_{de}(0)$ as a free function of redshift
and representing it by using the polynomial.
(iv) Gaussian Process modeling: using a distribution over functions that can represent $w_{de}(z)$ and estimating the statistical properties thereof.
These four classes of reconstruction methods and the related research works will be introduced in the following.

\subsubsection{Specific ansatz}

The ``specific ansatz'' is the most popular approach currently.
The key idea is assuming a specific parameterized form for $w_{de}(z)$ and estimating the associated parameters.
A simple and widely used ansatz is the XCDM ansatz, in which the EOS of dark energy is a constant, i.e. $w_{de} = const$.
This yields a simple form of $f(z)$
\begin{align}
\label{fXCDM}
f(z) = (1+z)^{3(1+w_{de})}.
\end{align}
In \cite{WMAPFive}, by combining the WMAP5 observations with BAO and SN data, Komatsu {\it et al.} obtained $w_{de}=-0.992_{-0.062}^{+0.061}$ at the 1$\sigma$ CL,
while in \cite{WMAPSeven}, a combination analysis of WMAP7+BAO+SN gave $w_{de}=-0.980_{-0.053}^{+0.053}$.
A more recent constraint on $w_{de}$ by the SCP team \cite{UnionTwo}\ also presented the consistent result.
So the current observations still favor $w_{de}=-1$ (i.e. cosmological constant).

Besides, one can also assume that the EOS of dark energy is not a constant.
The most popular parametrization with dynamical $w_{de}$ is the Chevallier-Polarski-Linder (CPL) parameterization \cite{CPLBE,CPL},
which assume $w_{de}(z) = w_{0} + w_{a}z/(1 + z)$.
The corresponding $f(z)$ is given by
\cite{CPLBE,CPLtwo}
\begin{align}
\label{fCPL}
f(z) = (1+z)^{3(1+w_0+w_a)}\exp\left(-{3w_{a}z\over1+z}\right).
\end{align}
Here $w_0$ denotes the value of the present EOS, while $w_a$ denotes the variation of the EOS.
The current observational data, such as the Union2 SNIa dataset \cite{UnionTwo} and the WMAP7 data \cite{WMAPSeven},
still favor the result of $w_0=-1$ and $w_a=0$, which is consistent with the cosmological constant.

In addition, some other ansatzs have also been proposed,
such as $w_{de}(z) = w_{0} + w_{1}z$ \cite{HutererTurner},
$w_{de}(z) = w_{0} + w_{a}z/(1 + z)^p$ \cite{JBP},
$w_{de}(z) = w_{0} + w_{1}\ln(1+z)$ \cite{Efstathiou},
and $w_{de}(z)=w_0+w_1(\frac{\ln(2+z)}{1+z}-\ln2)$ \cite{MZ}.
For more research works of various
parametrization forms, see \cite{DeParUSeljak,Perivolaropoulos,BABassett,Upadhye} and references therein.

\subsubsection{Binned parametrization}

In addition to the specific ansatz, another popular approach is the binned parametrization.
The binned parametrization was firstly proposed by Huterer and Starkman \cite{HutererStarkman}
based on the principal component analysis (PCA) \cite{HutererStarkman,HutererCooray}.
The key idea is dividing the redshift range into different bins and picking a simple local basis representation for $w_{de}(z)$.
The simplest way is setting $w_{de}(z)$ as piecewise constant in redshift,
thus $f(z)$ can be written as \cite{Sullivan,Qi}
\begin{align}
\label{fzwbinned}
f(z_{n-1}<z \le z_n)=(1+z)^{3(1+w_n)}\prod_{i=0}^{n-1}(1+z_i)^{3(w_i-w_{i+1})},
\end{align}
where $w_i$ is the EOS parameter in the i-th redshift bin defined by an upper boundary at $z_i$.

The optimal choice of redshift bins is still in debate.
In \cite{Goldsix}, Riess {\it et al.} proposed an uniform, unbiased binning method,
in which the number of SNIa in each bin times the width of each bin is a constant (i.e. $n \Delta z=const$).
In \cite{WangMPLA}, Wang argued that one should choose a constant $\Delta z$ for redshift slices,
to ensure the variation of $H$ and $1/D_A$ in each redshift slice remain roughly constant with $z$.
In \cite{FreeZione}, we presented a new binned parametrization method,
which treats $z_i$ as models parameters and let them run freely in the redshift region of SNIa samples.
This binning method can obtain smaller $\chi_{min}^{2}$ and tighter error bars \cite{FreeZitwo,FreeZithree}.
It should be mentioned that,
since the current observational data at high redshifts (i.e., z $>$ 1) are very rare,
for the binned parametrization, only two parameters of EOS $w_{de}$ can be constrained well \cite{CPLpar}.

In addition to the piecewise constant parametrization, some other
local basis representations for $w_{de}(z)$ are also
proposed, such as wavelet \cite{wavelet}\  and numerical derivatives
\cite{NCGasDaly,ShAlSaStr}.

\subsubsection{Polynomial fitting}

The third approach is the polynomial
fitting method. The key idea is treating the dark energy density function
$f(z)$ as a free function of redshift and representing it by using
the polynomial. Compared with the binned parametrization, the
advantage of the polynomial fitting parametrization is that the dark energy
density function $f(z)$ can be reconstructed as a continuous
function in the redshift range covered by the observational data.

A simple polynomial fit to $f(z)$ was proposed by Alam {\it et al.}
\cite{ASSSBefore}, which is a truncated Taylor expansion
\begin{align}
\label{Alam}
f(z)=A_0 + A_1(1+z) + A_2(1+z)^2.
\end{align}
This ansatz has only three free parameters ($\Omega_{m0},A_1,A_2$) since $A_0 + A_1 + A_2 = 1-\Omega_{m0}$ for a flat Universe.
Using this ansatz, Alam {\it et al.} \cite{ASSS} argued that the Tonry/Barris SNIa sample \cite{TonryBarris}
appear to favor dark energy which evolves in time
\footnote{It should be mentioned that there was a debate about the reliability of this ansatz \cite{Jonsson,ASSStwo}}.
Another interesting polynomial fit is the polynomial interpolation,
which was proposed by Wang
\cite{FOMWang}.
It choose different redshift points $z_i = i\ast z_{\rm max}/n (i = 1, 2,\cdots, n)$,
and interpolate $f(z)$ by using its own values at these
redshift points. This yields
\begin{align}
\label{YUN}
f(z)=\sum_{i=1}^{n}
f_i{(z-z_1)...(z-z_{i-1})(z-z_{i+1})...(z-z_n)\over(z_i-z_1)...(z_i-z_{i-1})(z_i-z_{i+1})...(z_i-z_n)}.
\end{align}
Here $f_i=f(z_i)$ and $z_n=z_{\rm max}$. Based on the relation $f(0)=1$,
one parameter can be fixed directly, and only $n-1$ model parameters need to be determined by the data.
For more research works of this polynomial interpolation method, see \cite{MoreWang}.

\subsubsection{Gaussian process modeling}

The fourth approach is the Gaussian Process (GP) modeling,\index{Gaussian Process (GP) modeling} which is proposed by Holsclaw {\it et al.} \cite{GPPRL,GPPRDOne}.
GP is a stochastic process indexed by $z$.
The defining property of a GP \index{Gaussian Process (GP)}
is that the vector that corresponds to the process at any finite collection of points follows a multivariate Gaussian distribution \cite{GPBaRa}.
GPs are elements of an infinite dimensional space, and can be used as the basis for a nonparametric reconstruction method.
They are characterized by a mean and a covariance function, defined by a small number of hyperparameters \cite{GPBaRa,GPPRDTwo}.

Based on the definition of a GP, one can assume that, for any collection $z_1$, ..., $z_n$,
$w_{de}(z_1)$, ..., $w_{de}(z_n)$ follows a multivariate Gaussian distribution with a constant negative mean and exponential covariance function written as
\begin{align}
\label{GPone}
K(z, z') = \kappa^2 x^{|z-z'|^{\alpha}}.
\end{align}
The hyperparameters $x \in (0, 1)$ and $\kappa$, and the parameters defining the likelihood, are determined by the data.
The value of $\alpha \in (0, 2]$ influences the smoothness of the GP realizations:
for $\alpha=2$, the realizations are smooth with infinitely many derivatives,
while $\alpha=1$ leads to rougher realizations suited to modeling continuous non-(mean-squared)-differentiable functions.
Moreover, one can set up the following GP for $w_{de}$
\begin{align}
\label{GPtwo}
w_{de}(u) \sim GP(-1, K(u, u')).
\end{align}
Making use of the Eq. (\ref{GPtwo}),
one can take advantage of the particular integral structure of luminosity distance $d_L(z)$ expressed by $w_{de}(z)$ (see \cite{GPPRL,GPPRDOne} for details).

\section{Concluding Remarks}

We have reviewed theoretical models as well as observational and numerical studies of dark energy.
Numerous works and papers have been done and written since the discovery of the accelerating expansion of the Universe,
it is impossible to cover even a small part of the heroic endeavors of our community in any review article.

However, the problem of understanding the nature of dark energy is as daunting as ever,
or perhaps some already hold the key to this understanding without being commonly accepted yet.
Clearly, there is a long long way to go for both theorists and experimentalists.

It is without any doubt that the process of detecting the nature of dark energy and understanding
its origin will prove to be one of the most exciting stories in modern science.

\section*{Acknowledgment}
This research was supported by a NSFC grant No.10535060/A050207, a NSFC grant No.10975172, a NSFC group grant No.10821504 and Ministry of Science and Technology 973 program under grant No.2007CB815401. YW acknowledges support from grants from Kavli IPMU, McGill University, the Institute of Particle Physics (Canada) and the Foundational Questions Institute.

\end{document}